\documentclass{hep99}

\usepackage{cite}
\usepackage{epsfig}

\def\beq{\begin{equation}}
\def\eeq{\end{equation}}
\def\beqar{\begin{eqnarray}}
\def\eeqar{\end{eqnarray}}
\def\barr#1{\begin{array}{#1}}
\def\earr{\end{array}}
\def\bfi{\begin{figure}}
\def\efi{\end{figure}}
\def\btab{\begin{table}}
\def\etab{\end{table}}
\def\bce{\begin{center}}
\def\ece{\end{center}}

\def\text{\textstyle}

\def\Ga{\Gamma}

\def\refeq#1{\mbox{eq.~(\ref{#1})}}
\def\refeqs#1{\mbox{eqs.~(\ref{#1})}}
\def\reffi#1{\mbox{Fig.~\ref{#1}}}

\def\citere#1{\mbox{Ref.~\cite{#1}}}

\def\mathswitchr#1{\relax\ifmmode{\mathrm{#1}}\else$\mathrm{#1}$\fi}

\newcommand{\PW}{\mathswitchr W}

\newcommand{\PA}{\mathswitchr A}

\newcommand{\PH}{\mathswitchr H}

\newcommand{\Ph}{\mathswitchr h}

\newcommand{\Pb}{\mathswitchr b}

\newcommand{\Pt}{\mathswitchr t}

\def\mathswitch#1{\relax\ifmmode#1\else$#1$\fi}

\newcommand{\MW}{\mathswitch {M_\PW}}

\newcommand{\Mt}{\mathswitch {m_\Pt}}
\newcommand{\Mh}{\mathswitch {m_\Ph}}
\newcommand{\mh}{\mathswitch {m_\Ph}}

\newcommand{\MA}{\mathswitch {M_\PA}}

\newcommand{\GF}{\mathswitch {G_\mu}}

\def\tb{\tan\beta}

\newcommand{\MstL}{M_{\tilde{t}_L}}
\newcommand{\MstR}{M_{\tilde{t}_R}}

\newcommand{\At}{A_{\Pt}}
\newcommand{\Ab}{A_{\Pb}}
\newcommand{\Xt}{X_{\Pt}}

\newcommand{\mt}{\Mt}

\newcommand{\mgl}{m_{\tilde{\mathrm{g}}}}

\newcommand{\tsf}{\theta\kern-.20em_{\tilde{f}}}
\newcommand{\tsfp}{\theta\kern-.20em_{\tilde{f}\prime}}
\newcommand{\tsq}{\theta\kern-.15em_{\tilde{q}}}

\newcommand{\msusy}{M_{\mathrm{SUSY}}}

\newcommand{\lsim}
{\;\raisebox{-.3em}{$\stackrel{\displaystyle <}{\sim}$}\;}

\newcommand{\alps}{\alpha_{\mathrm s}}

\newcommand{\fh}{{\em FeynHiggs}}

\newcommand{\subh}{{\em subhpole}}

\newcommand{\msbar}{$\overline{\rm{MS}}$}

\newcommand{\oaas}{{\cal O}(\alpha\alps)}
\newcommand{\cp}{{\cal CP}}

\newcommand{\twol}{two-loop}
\newcommand{\onel}{one-loop}

\newcommand{\mhmax}{\mh^{\mathrm{max}}}

\newcommand{\KL}{\left(}
\newcommand{\KR}{\right)}

\newcommand{\VL}{\left( \begin{array}{c}}
\newcommand{\VR}{\end{array} \right)}
\newcommand{\ML}{\left( \begin{array}{cc}}
\newcommand{\MLd}{\left( \begin{array}{ccc}}
\newcommand{\MLv}{\left( \begin{array}{cccc}}
\newcommand{\MR}{\end{array} \right)}

\newcommand{\tev}{\,\, \mathrm{TeV}}
\newcommand{\gev}{\,\, \mathrm{GeV}}

\newcommand{\BC}{\begin{center}}
\newcommand{\EC}{\end{center}}
\newcommand{\BE}{\begin{equation}}
\newcommand{\EE}{\end{equation}}
\newcommand{\BEA}{\begin{eqnarray}}
\newcommand{\BEAnn}{\begin{eqnarray*}}
\newcommand{\EEA}{\end{eqnarray}}
\newcommand{\EEAnn}{\end{eqnarray*}}
\newcommand{\non}{\nonumber}
\newcommand{\id}{{\rm 1\kern-.12em
\rule{0.3pt}{1.5ex}\raisebox{0.0ex}{\rule{0.1em}{0.3pt}}}}

\def\hSi{\hat{\Sigma}}

\def\draftdate{\relax}
\def\mda{\relax}
\def\mua{\relax}
\def\mla{\relax}
\def\draft{
\def\thtystars{******************************}
\def\sixtystars{\thtystars\thtystars}
\typeout{}
\typeout{\sixtystars**}
\typeout{* Draft mode!
         For final version remove \protect\draft\space in source file
*}
\typeout{\sixtystars**}
\typeout{}
\def\draftdate{\today}
\def\mua{\marginpar[\boldmath\hfil$\uparrow$]%
                   {\boldmath$\uparrow$\hfil}%
                    \typeout{marginpar: $\uparrow$}\ignorespaces}
\def\mda{\marginpar[\boldmath\hfil$\downarrow$]%
                   {\boldmath$\downarrow$\hfil}%
                    \typeout{marginpar: $\downarrow$}\ignorespaces}
\def\mla{\marginpar[\boldmath\hfil$\rightarrow$]%
                   {\boldmath$\leftarrow $\hfil}%
                    \typeout{marginpar:
$\leftrightarrow$}\ignorespaces}
\def\Mua{\marginpar[\boldmath\hfil$\Uparrow$]%
                   {\boldmath$\Uparrow$\hfil}%
                    \typeout{marginpar: $\Uparrow$}\ignorespaces}
\def\Mda{\marginpar[\boldmath\hfil$\Downarrow$]%
                   {\boldmath$\Downarrow$\hfil}%
                    \typeout{marginpar: $\Downarrow$}\ignorespaces}
\def\Mla{\marginpar[\boldmath\hfil$\Rightarrow$]%
                   {\boldmath$\Leftarrow $\hfil}%
                    \typeout{marginpar:
$\Leftrightarrow$}\ignorespaces}
\overfullrule 5pt
\oddsidemargin -15mm
\marginparwidth 29mm
}

\begin{document}

\onecolumn
\thispagestyle{empty}
\setcounter{page}{0}
\def\thefootnote{\fnsymbol{footnote}}

\begin{flushright}
CERN--TH/99--367\\
DESY 99--183\\
KA--TP--23--1999\\
hep-ph/9912263 \\
\date{\today}
\end{flushright}

\vspace{1cm}

\begin{center}

{\large\sc {\bf Higher-order results in the Higgs sector }}

\vspace*{0.4cm} 

{\large\sc {\bf of the MSSM}}
\footnote{Presented by G.~Weiglein at the ``International Europhysics
  Conference on High Energy Physics'', Tampere, July 1999.}

\vspace{1cm}

{\sc S.~Heinemeyer$^{1\,}$%
\footnote{
email: Sven.Heinemeyer@desy.de
}%
, W.~Hollik$^{\,2}$%
\footnote{
email: Wolfgang.Hollik@physik.uni-karlsruhe.de
}%
 and G.~Weiglein$^{\,3}$%
\footnote{
email: Georg.Weiglein@cern.ch
}%
}

\vspace*{1cm}

$^1$ DESY Theorie, Notkestr.\ 85, D--22603 Hamburg, Germany

\vspace*{0.4cm}

$^2$ Institut f\"ur theoretische Physik, Universit\"at Karlsruhe,
D--76128 Karlsruhe, Germany

\vspace*{0.4cm}

$^3$ CERN, TH Division, CH--1211 Geneva 23, Switzerland

\end{center}

\vspace*{1cm}

\BC
{\bf Abstract}
\EC
We analyze the impact of the recent Feynman-diagrammatic (FD) 
two-loop results for the mass of the lightest $\cp$-even
Higgs boson in the MSSM on the theoretical upper bound for $\mh$ as a 
function of $\tan\beta$. The 
results are compared with previous results obtained by renormalization 
group (RG) methods. 
The incorporation of dominant FD two-loop corrections into the 
decay width $\Ga(h \to f \bar f)$ is also discussed.
%

\def\thefootnote{\arabic{footnote}}
\setcounter{footnote}{0}

\newpage
\twocolumn


\title{Higher-order results in the Higgs sector of the MSSM}

\author{S.~Heinemeyer$^1$, W.~Hollik$^2$ and G.~Weiglein$^3$}
\address{$^1$ DESY Theorie, Notkestr.\ 85, D--22603 Hamburg, Germany\\
$^2$ Institut f\"ur theoretische Physik, Universit\"at Karlsruhe,
D--76128 Karlsruhe, Germany\\
$^3$ CERN, TH Division, CH--1211 Geneva 23, Switzerland\\[3pt]
E-mails: {\tt Sven.Heinemeyer@desy.de,
Wolfgang.Hollik@physik.uni-karlsruhe.de, Georg.Weiglein@cern.ch}}

\abstract{
We analyze the impact of the recent Feynman-diagrammatic (FD) 
two-loop results for the mass of the lightest $\cp$-even
Higgs boson in the MSSM on the theoretical upper bound for $\mh$ as a 
function of $\tan\beta$. The 
results are compared with previous results obtained by renormalization 
group (RG) methods. 
The incorporation of dominant FD two-loop corrections into the 
decay width $\Ga(h \to f \bar f)$ is also discussed.
}

\maketitle

\section{Introduction}

The lightest $\cp$-even Higgs boson in the Minimal
Supersymmetric Standard Model (MSSM) is of particular interest, since it
is bounded to be lighter than the Z~boson at the tree level. The \onel\
results~\cite{mhiggs1l,mhiggsf1l,mhiggsf1lb} for its mass, $\mh$, have
in the last years been extended by the leading \twol\ corrections,
performed in the renormalization group (RG) approach~\cite{mhiggsRG}, in
the effective potential approach~\cite{mhiggsEP} and most recently in 
the Feynman-diagrammatic (FD) approach~\cite{mhiggsFD}. These calculations 
predict an upper bound on $\mh$ of about $\mh \lsim 135 \gev$.

A precise prediction for the mass of the lightest $\cp$-even Higgs boson
as well as for the cross sections of its production and decay processes
is important for the Higgs-boson search at LEP2, the upgraded Tevatron and the
LHC. If the lightest $\cp$-even Higgs boson will be found at the present
or the next generation of colliders, its mass will be determined with a
high precision, allowing thus a sensitive test of the model. 

The dominant radiative corrections to $\mh$ arise from the top and
scalar top sector of the MSSM, with the input parameters $\mt$, $\msusy$ and 
$\Xt$. For simplicity, the soft SUSY breaking parameters in the diagonal 
entries of the scalar top mixing matrix are often assumed to be equal,
$\msusy = \MstL = \MstR$. The off-diagonal entry of the mixing
matrix in our conventions (see \citere{mhiggsFD}) reads 
$\mt \Xt = \mt (A_{\Pt} - \mu \cot\beta)$.

Up to now most phenomenological analyses have been based on the results
obtained within the RG approach~\cite{mhiggsRG}, where the neutral 
$\cp$-even Higgs-boson masses are calculated from the effective
couplings in the Higgs potential. The results contain the
leading logarithmic contributions at the two-loop level.

In the FD approach, on the other hand, the masses of the $\cp$-even
Higgs bosons are obtained by evaluating loop corrections to the 
$h$, $H$ and $hH$-mixing propagators and by determining the poles of the
corresponding propagator matrix.
In \citere{mhiggsFD} the dominant two-loop
contributions to the masses of the $\cp$-even Higgs bosons of $\oaas$ 
have been evaluated in the on-shell renormalization scheme. They have 
been combined with the complete one-loop on-shell result~\cite{mhiggsf1lb} 
and the sub-dominant two-loop corrections of 
${\cal O}(\GF^2 \mt^6)$~\cite{mhiggsRG}. The corresponding results
have been implemented into the Fortran code \fh~\cite{feynhiggs}.

In \citere{mhiggslle} the dominant contributions have been extracted
from the full FD result. Taking into account the fact that the FD and
the RG result have been obtained within different renormalization
schemes and transforming the FD result of \citere{mhiggslle} into the 
\msbar\ scheme, it has been shown that the RG and the FD 
approach agree in the leading logarithmic terms at the two-loop 
level~\cite{bse}. 
The FD result, however, contains further genuine two-loop terms of 
non-logarithmic nature that go beyond the RG result. These genuine two-loop 
terms lead to an increase of the maximal value of $\mh$ compared to the 
RG result of up to $5 \gev$~\cite{bse,prec}.



\section{Implications for $\tb$ exclusion limits}

By combining the theoretical result for the upper bound on $\mh$ as a
function of $\tb$ in the MSSM with the informations from the direct search
for the lightest Higgs boson, it is possible to derive constraints on
$\tb$.
Since the predicted value of $\mh$ depends
sensitively on the precise numerical value of $\mt$, it has become
customary to discuss the constraints on $\tb$ within a so-called
``benchmark'' scenario, in which $\mt$ is kept fixed at the value $\mt =
175 \gev$ and in which furthermore a large value of $\msusy$ is chosen,
$\msusy = 1 \tev$, giving rise to large values of $\mh(\tb)$.

In \citere{tbexcl} it has recently been analyzed how the values chosen
for the other SUSY parameters in the benchmark scenario 
(see \citere{Higgsgroup} and references therein) should be modified
in order to obtain the maximal values of $\mh(\tb)$ for given $\mt$ and 
$\msusy$. The maximal values for $\mh$ as a function of $\tb$ within this
scenario ($\mhmax$ scenario) are higher by about 5~GeV than in the
usual benchmark scenario. The constraints on $\tb$ derived within the
$\mhmax$ scenario are thus more conservative than the ones based on the
previous benchmark scenario.

The $\mhmax$ scenario is defined as~\cite{tbexcl,bench}
\BEA
{} \hspace{-2.5em}  && \mt = \mt^{\mathrm{exp}}~  ( = 174.3 \gev ), \,
        \msusy = 1 \tev \non \\
{} \hspace{-2.5em}  && \mu = -200 \gev,\,
        M_2 = 200 \gev,  \, \MA \leq 1000 \gev \non \\
{} \hspace{-2.5em} && \Xt = 2 \, \msusy \quad \mbox{(FD calculation)} \non \\ 
{} \hspace{-2.5em} && \Xt = \sqrt{6} \, \msusy \quad 
             \mbox{(RG calculation)} \non \\ 
{} \hspace{-2.5em} && \mgl = 0.8 \, \msusy \quad \mbox{(FD calculation),} 
\label{newbenchmarkdef}
\EEA
where the parameters are chosen such that the chargino masses are beyond
the reach of LEP2. In \refeq{newbenchmarkdef} 
$\mu$ is the Higgs mixing parameter, $M_2$ denotes the
soft SUSY breaking parameter in the gaugino sector, and $\MA$ is 
the $\cp$-odd Higgs-boson mass.

Different values of $\Xt$ are specified in \refeq{newbenchmarkdef} for
the results of the FD and the RG calculation, since within the two
approaches the maximal values for $\mh$ are obtained for different
values of $\Xt$. This fact is partly due to the different
renormalization schemes used in the two approaches, i.e.\ the parameter
$\Xt$ in the \msbar\ scheme is shifted with respect to the
corresponding parameter in the on-shell scheme~\cite{bse,prec}.
In \fh\ the gluino mass, $\mgl$, can be specified as a free
input parameter. The effect of varying $\mgl$ on $\mh$ is up to 
$\pm 2 \gev$~\cite{mhiggsFD}. Within the RG result~\cite{mhiggsRG} used
so far for the analysis of the benchmark scenario, $\mgl$ is 
fixed to $\mgl = \msusy$. 
The corresponding values of $\mh$ are about $0.5 \gev$ lower than the 
maximal values (obtained for $\mgl \approx 0.8\,\msusy$).

While so far we have only been concerned with the definition of an
appropriate scenario, we now turn to
the impact of the new FD two-loop result for $\mh$, which contains
previously unknown non-logarithmic two-loop terms. Comparing the FD
result (program \fh) with the RG result (program \subh, based on the
second and third reference of~\cite{mhiggsRG}) we find that the maximal
value for $\mh$ within the FD result is higher by up to 4~GeV.  

\begin{figure}
\begin{center}
\mbox{
\psfig{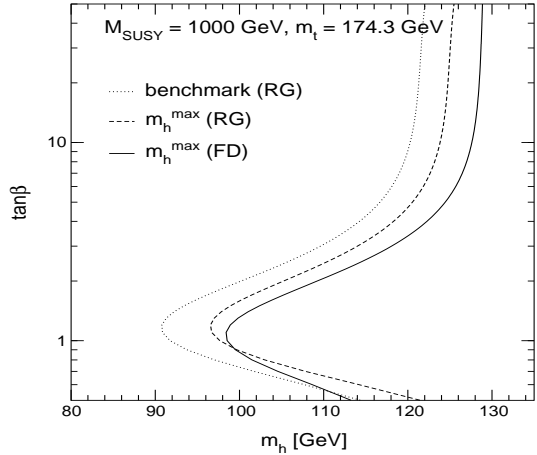}}
\end{center}
\caption[]{$\mh$ is shown as a function of $\tb$.
The dashed curve displays the RG result within the benchmark scenario,
while the dotted curve shows the RG result for the $\mhmax$ scenario
(program \subh). The solid curve corresponds to the FD result in the
$\mhmax$ scenario (program \fh).
}
\label{fig:mhtb}
\end{figure}

In \reffi{fig:mhtb} we show both the effect of modifying the previous
benchmark scenario to the $\mhmax$ scenario and the impact of the new FD
two-loop result on the prediction for $\mh$. The Higgs-boson mass is
plotted as a function of $\tb$. The dashed curve displays the 
previous benchmark scenario, while the dotted curve shows the $\mhmax$
scenario. Both curves are based on the RG result (program \subh). 
The full curve corresponds to the FD result (program \fh) in the 
$\mhmax$ scenario. The increase in the maximal value for $\mh$ by about
$4$~GeV from the new FD result and by further 5~GeV if the benchmark
scenario is replaced by the $\mhmax$ scenario has a significant effect
on exclusion limits for $\tb$ derived from the Higgs-boson search.
Using the FD result for the $\mhmax$ scenario,
an excluded $\tb$ region appears only for an experimental bound on 
$\mh$ of roughly 100~GeV, while in the previous benchmark definition 
values up to $\tb = 2$ would already be excluded for a Higgs limit of about
100~GeV.

The constraints on $\tb$ discussed above are of course only valid under
the specific assumptions made in the 
considered scenario. 
In particular, increasing $\mt$ by one or even two standard deviations
above the current experimental central value leads to a significant
increase in the maximal value of $\mh(\tb)$; 
increasing $\mt$ 
by 1~GeV roughly translates into an upward shift of $\mhmax$ of 1~GeV.



\section{
$\oaas$ Yukawa contributions to the decay width 
$\Ga(h \to f \bar f)$}

As an extension of the FD two-loop results for the neutral $\cp$-even
Higgs-boson masses, we consider now the leading two-loop 
Yukawa
corrections of ${\cal O}(\GF \alps\mt^4/\MW^2)$ to the decay width 
$\Ga(h \to f \bar f)$. These contributions enter the decay amplitude
$A(h \to f \bar f)$ in the following way,
\beq
A(h \to f \bar f) = \sqrt{Z_\Ph} (\Ga_\Ph + Z_{\Ph\PH} \Ga_\PH).
\label{eq:ahff}
\eeq
$\Ga_\Ph$, $\Ga_\PH$ are the $hf \bar f$ and $Hf \bar f$ vertex
functions, and
\BEA
Z_\Ph &\hspace{-.5em} = \hspace{-.5em} & \frac{1}{1 + \hSi'_{\Ph}(q^2) -
        \KL \frac{\hSi_{\Ph\PH}^2(q^2)}
       {q^2 - m_{\PH, (0)}^2 + \hSi_\PH(q^2)} \KR '}
  \Biggr|_{q^2 = \mh^2}, \non \\
Z_{\Ph\PH} & \hspace{-.5em} = \hspace{-.5em} & - \frac{\hSi_{\Ph\PH}(\mh^2)}
                {\mh^2 - m_{\PH, (0)}^2 + \hSi_\PH(\mh^2)} .
\label{eq:ZhH}
\EEA
Here $\hSi_{\Ph}(q^2)$, $\hSi_{\Ph\PH}(q^2)$,
$\hSi_{\PH}(q^2)$ denote the real parts of the renormalized Higgs-boson
self-energies and $m_{\PH, (0)}$ is the tree-level mass of the heavier 
$\cp$-even Higgs boson.

\begin{figure}
\begin{center}
\mbox{
\psfig{figure=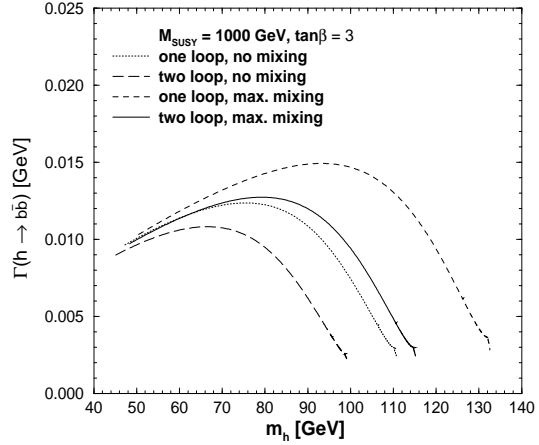,width=7cm,height=6cm}}
\end{center}
\caption[]{
The results for $\Ga(h\to b\bar b)$ containing the Higgs propagator corrections
at one- and two-loop order are shown as a function of $\Mh$. The results
are given in the no-mixing and maximal-mixing case 
for $\mu = -100 \gev$, $M_2 = \msusy$, $\mgl = 500 \gev$, 
$\Ab = \At$.
}
\label{fig:Ghbb}
\end{figure}

In \reffi{fig:Ghbb} the results for the decay width $\Ga(h \to b \bar b)$ 
including the two-loop propagator corrections according to
\refeqs{eq:ahff}--(\ref{eq:ZhH}) are compared with the corresponding
one-loop result for the cases of no mixing and maximal mixing in the
scalar top sector. In both results the one-loop QED and QCD (gluon and
gluino exchange) vertex corrections are included. The effect of
the two-loop contributions is seen to be sizable. Since the branching
ratio $\mbox{BR}(h \to b \bar b)$ is in general strongly dominated by 
$\Ga(h \to b \bar b)$, the correction to a large extent cancels out in
the branching ratio. A comparison of our FD results with the
corresponding results obtained within the RG approach is given in
\citere{hff}.



\end{document}